\documentclass[aps,prb,preprint,groupedaddress,amsmath,amssymb]{revtex4-1}
\usepackage{times,xspace}
\usepackage{graphicx,graphics,color,epsfig}
\usepackage{amsmath}
\usepackage{amssymb}
\usepackage{amsfonts}
\usepackage{amsfonts}
\usepackage{epstopdf}
\usepackage{bm}
\usepackage{color}

\begin{document}


\title{Theory of nodal $s^{\pm}$-wave pairing symmetry in the Pu-based 115 superconductor family}

\author{Tanmoy Das$^{1,*}$, Jian-Xin~Zhu$^{1,2}$, \& Matthias J.~Graf$^{1}$}
\affiliation{$^1$Theoretical Division, Los Alamos National Laboratory, Los Alamos, NM 87545, USA.}
\affiliation{$^2$Center for Integrated Nanotechnologies, Los Alamos National Laboratory, Los Alamos, NM 87545, USA.}
\affiliation{$^*$Correspondence should be made to tnmydas@gmail.com}

\date{\today}

\begin{abstract}
The spin-fluctuation mechanism of superconductivity usually results in the presence of gapless or nodal quasiparticle states in the excitation spectrum. Nodal quasiparticle states are well established in copper-oxide, and heavy-fermion superconductors, but not in iron-based superconductors. Here, we study the pairing symmetry and mechanism of a new class of plutonium-based high-Tc superconductors and predict the presence of a nodal $s^{+-}$ wave pairing symmetry in this family. Starting from a density-functional theory (DFT) based electronic structure calculation we predict several three-dimensional (3D) Fermi surfaces in this 115 superconductor family. We identify the dominant Fermi surface ``hot-spots' in the inter-band scattering channel, which are aligned along the wavevector ${\bf Q}\sim(\pi,\pi,\pi)$, where degeneracy could induce sign-reversal of the pairing symmetry. Our calculation demonstrates that the $s^{+-}$ wave pairing strength is stronger than the previously thought $d$-wave pairing; and more importantly, this pairing state allows for the existence of nodal quasiparticles. Finally, we predict the shape of the momentum- and energy-dependent magnetic resonance spectrum for the identification of this pairing symmetry.
\end{abstract}

\maketitle

The unconventional mechanism of Cooper pairing is often attributed to a momentum dependent superconducting (SC) gap structure with sign reversal of its amplitude, which renders the electron pair formation in a repulsive potential background~\cite{scalapino}. This scenario is primarily supported by the proximity of superconductivity to magnetism, and a correspondence between the energy of the magnetic resonance mode and the SC transition temperature ($T_c$)~\cite{Greven}. The recently discovered $5f$-based intermetallic actinides Pu$M$$T_5$ ($M=$Co, Rh, and $T=$In, Ga) or in short Pu-115 represent a rather exotic class of superconductors, which exhibit evidence of nodal superconductivity despite the absence of a magnetic instability. However, the nuclear magnetic response (NMR)~\cite{Curro,Sakai,Baek} and muon spin rotation ($\mu$SR)
measurements~\cite{muSR1,muSR2}, later on supported by theoretical study~\cite{DasPRL}, demonstrate that spin fluctuations are significantly large, but  do not drive the system into a  magnetic ground state at low temperatures~\cite{muSR1}. Therefore, a novel materials specific theoretical route to link spin fluctuations to superconductivity would be helpful in order to test whether there is a universal magnetic mechanism of superconductivity in all families of  high-$T_c$ superconductors.\cite{JCDavis}

The intermetallic Pu-115 actinide superconductors represent the link in the sequence going from the heavy-fermions ($T_c\sim$1~K), to the iron-pnictides ($T_c\sim$50~K), on to the cuprates ($T_c\sim$100~K), with respect to the values of the SC transition temperature $T_c$, spin fluctuation temperature $T_s$, as well as mass enhancement~\cite{Curro,Sakai}.
PuCoGa$_5$ possesses an electron mass renormalization of $m^*/m_b\sim 3.5$ (where $m_b$ is the DFT-deduced band mass), which is many times smaller than that of the heavy-fermion superconductor CeCoIn$_5$ with $m^*/m_b\sim 30$~\cite{Movshovich,Settai}. Such a moderate mass enhancement by electron correlations is well captured by dressing of $5f$-states via the spin-fluctuation coupling of electron-hole excitations~\cite{DasPRL,DasPRX}.
The exchange of spin fluctuations divides the electronic states into renormalized itinerant quasiparticles near the Fermi level and strongly localized incoherent states at higher binding energies. This scenario of dynamical correlation effects is consistent with the corresponding peak-dip-hump feature measured in photoemission spectroscopy~\cite{PESPucoGa5}. The bulk Curie-Weiss susceptibility observed in the normal state of
PuCoGa$_5$~\cite{PuCoGa5} was initially interpreted to arise from static moments~\cite{muSR1,PNeutron} but more recently suggested to come from the fluctuation of spins or valences of Pu between $5f^5$ and $5f^{6}$ configurations of the ground state~\cite{Pezzoli,Shick2013}, as in some heavy-fermion superconductors~\cite{Miyake07}. Taken together, Pu-115 compounds reside in between localized and itinerant systems,  for which the DFT band structure with proper mass renormalization may be the appropriate starting point to describe the low-energy spectrum, from which superconductivity emerges.

Our main finding of the DFT-based calculations of the pairing symmetry, due to a Fermi surface (FS) instability, is that a nodal $s^{\pm}$-wave pairing symmetry is favored over the previously thought $d_{x^2-y^2}$-wave pairing symmetry. Despite the presence of four different FS pieces, the leading pairing instability arises from the enhanced scattering between hole pockets at the $\Gamma$ point and electron pockets at the A=$(\pi,\pi,\pi)$ point in the Brillouin zone. This FS topology acquires an analogy to the electronic states of the more recently discovered iron-based superconductors. However, unlike in the latter family, here the $s^{\pm}$-wave pairing is significantly anisotropic due to nearest-neighbor electron pairing, and its nodal planes intercept the hole-like FS near the zone boundary. We further present DFT-based results of the magnetic excitation spectrum for both pairing symmetries, and observe a prominent collective spin-1 mode, which is localized in both energy and momentum within the SC phase for the $s^{\pm}$-wave rather than for the $d$-wave symmetry. Our result of a nodal pairing state is consistent with power-law signatures in the spin-lattice relaxation rate~\cite{Curro,Sakai,Baek}, superfluid density~\cite{muSR1,muSR2}, and zero-bias conductance peak in point-contact spectroscopy (PCS)~\cite{PCT}, among others. Although these results are often taken to be consistent with the assumption of a nodal $d$-wave pairing, it should be noted that these probes are only sensitive to the nodal states, not to their location on the FS. To clarify this issue we carry out a multiband PCS calculation using both nodal pairing symmetries. We find that a zero-bias conductance peak is generated in both cases, and that the PCS experimental data\cite{PCT} can well be reproduced by the nodal $s^{\pm}$ pairing. Finally, the spin-fluctuation mediated pairing symmetry study provides a microscopic explanation of the pairing mechanism and order parameter symmetry in this less-explored plutonium-based family of superconductors. Such a study is needed because the existence of the sign-reversal $s^{\pm}$-pairing symmetry without a node can be difficult to distinguish from the $s^{++}$ symmetry, as is the case in the iron pnictides \cite{Kontani}, unless the sign of the order parameter can be directly measured. On the other hand, the existence of nodes in the 115 compounds may help establish that an unusual nodal $s^{\pm}$-pairing symmetry indeed exists.

\begin{figure*}[h]
\centering
\rotatebox[origin=c]{0}{\includegraphics[width=0.98\textwidth]{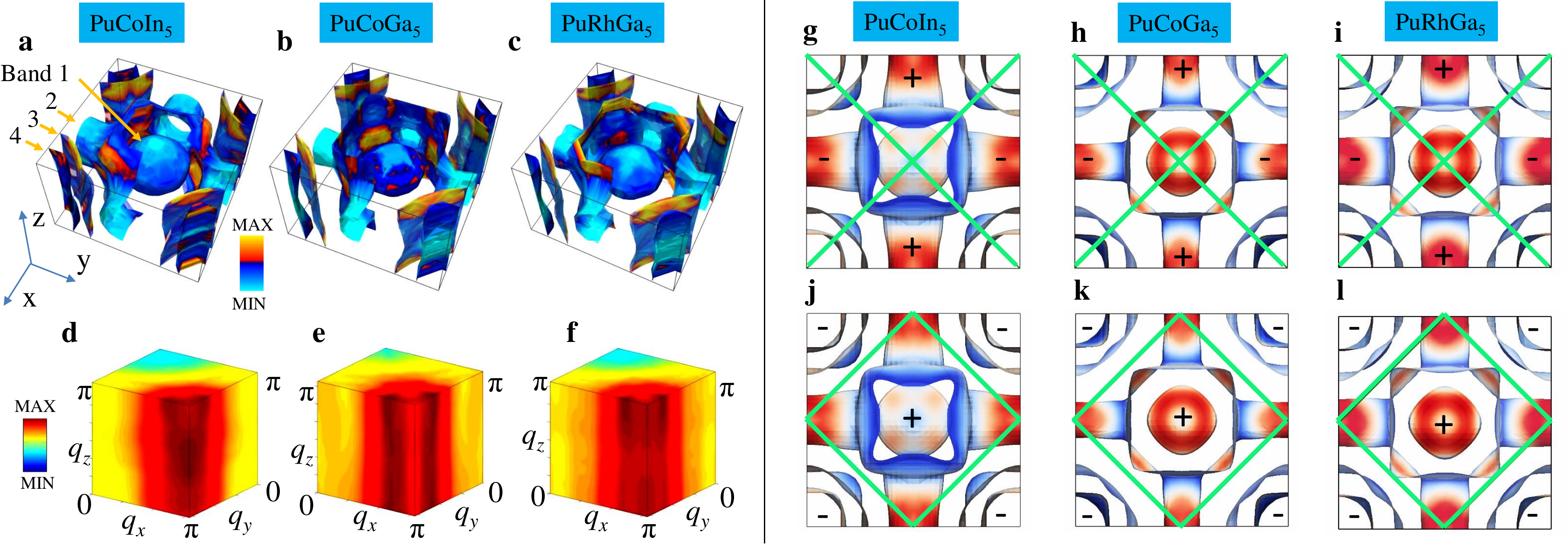}}
\caption{{Fermi surfaces and {\it hot-spots} from first-principles electronic band structure calculations.}
(a)-(c) FS topologies in the momentum space of the Brillouin zone for all three Pu-115 materials, plotted separately in three columns. An intensity colormap is used to depict the values of the JDOS on the FSs at ${\bf Q}=(\pi,\pi,\pi)$. For better visualization a quadrant of the holelike FS of band 2 was clipped.
The JDOS gives a qualitative estimate of the static susceptibility (Eq.~(\ref{eq:sus}))  for nesting vector ${\bf Q}$. These images help identify the strongest nesting between bands 1 and 2 (in the vicinity of the $k_z=0$-plane) to bands 3 and 4 (near the $k_z=\pi$-plane).
(d)-(f)
The full momentum (${\bf q}$) dependence of the static bare bubble susceptibility $\chi({\bf q}, \omega=0)$ is visualized in three-dimensional volume rendering. The highest intensity (red color) is in the vicinity of ${\bf q} \sim (\pi,\pi,q_z)$.
(g)-(i)
Top views of FSs (same as in Fig.~1a-c) with corresponding colormaps of the magnitude of the Fermi velocities (or inverse normal-state density of states) from low (blue) to high (red).
The green solid lines denote the nodal planes of the SC $d_{x^2-y^2}$-wave pairing symmetry.
(j)-(l)
Same as above, but now the green solid lines denote the nodal planes of the SC $s^\pm$-wave pairing symmetry.
Note, only the holelike FS of band 2 has nodes in the gap function on the cross-arms near the zone boundary.
} \label{fig1}
\end{figure*}

\vskip12pt
\noindent
{\bf Results}\\
{\bf Fermi Surface Nesting and Hot-Spots.} We begin by evaluating the nature of enhanced FS scattering or {\it hot-spots}, and the electronic fingerprints of $s^{\pm}$-, and $d_{x^2-y^2}$-wave pairing symmetries for three known Pu-115 superconductors PuCoIn$_5$ ($T_c$=2.5~K)~\cite{PuCoIn5}, PuCoGa$_5$ ($T_c$=18.5~K)~\cite{PuCoGa5}, and PuRhGa$_5$ ($T_c$=8~K)~\cite{PuRhGa5}. The low-energy electronic states of these compounds consist of four pairs of spin-orbit split energy bands cut by the Fermi level, as shown in Fig.~\ref{fig1}{a-c}.~\cite{DasPRL,JXZhu2012} We note that these results are in agreement with similar electronic structure calculations performed independently by other groups~\cite{Shick2013,Oppeneer,Opahle,Shick2011}. We estimate the strength of the band-dependent scattering enhancement by computing the bare bubble two-particle response function from first-principles band structure as
\begin{equation}
\chi_{nm}({\bf q},\omega)=-\sum_{\bf k}~\frac{f(\xi^n_{{\bf k}+{\bf q}}) - f(\xi^m_{{\bf k}})}{\omega + i\delta + \xi^n_{{\bf k}+{\bf q}} - \xi^m_{{\bf k}}},
\label{eq:sus}
\end{equation}
where $\xi^n_{\bf k}$ is the DFT-derived Bloch dispersion with wavevector ${\bf k}$ and band index $n$,
and $f(\xi_{\bf k}^{n})$ is the corresponding fermion occupation number. Figure~\ref{fig1}{d-f} shows the computed static susceptibilities in a colormap plot in the three-dimensional momentum transfer ${\bf q}$ space. 
The location of the maximum of $\chi({\bf q}) = \sum_{n,m}\chi_{nm}({\bf q}, 0)$ is primarily in the vicinity of ${\bf Q}\sim(\pi,\pi,\pi)$, with additional weights spread all along $q_z$. This suggests that the dominant FS instability occurs between the FSs separated by ${\bf Q}$ in the Brillouin zone. For this value of ${\bf Q}$, we identify the locations of the electronic  hot-spots or the highest joint-density of states (JDOS), which satisfy ${\bf Q}={\bf k}_i^n-{\bf k}_f^m$, where ${\bf k}_i^n$ and ${\bf k}_f^m$ are the Fermi momenta in the initial and final states of bands $n$ and $m$, respectively. The hot-spots are superimposed on the FSs using an intensity colormap as shown in Figs.~\ref{fig1}{a-c}. The intensity is determined from the approximate JDOS $\sim \frac{1}{v_{{\bf k},i}^n} \frac{1}{v_{{\bf k},f}^m} \delta_{{\bf Q}, {\bf k}_i^n-{\bf k}_f^m}$, where $v_{\bf k}^m$ is the Fermi speed in band $m$. We immediately see a consistent scenario for all three materials, namely that the hot-spots connect bands 1 or 2 near the plane with $k_z=0$ to bands 3 or 4 lying in the plane with $k_z=\pm\pi$. The locations of the hot-spots dictate a pairing symmetry, which favors sign reversal for ${\bf Q}$. In Fig.~\ref{fig1}{g-l} the same FS topologies are shown in top view with nodal lines for $d_{x^2-y^2}$-wave (top row) and $s^{\pm}$-wave (bottom row) pairing symmetries superimposed by green solid lines. Based on the correspondence between topology of the FSs and the dominant hot-spot nesting vector, it is now possible to conjecture that the Pu-115 system may favor  $s^{\pm}$-wave pairing. Wang et al.~\cite{Wang06} attained very similar results for $\chi_{nm}({\bf q},0)$, however, they emphasized the nesting at ${\bf Q}\sim(\pi,\pi,0)$.

\vskip12pt
\noindent
{\bf Electron Dispersions and Density of States.} Next we delineate the origin of the nodal state and compare the associated nodal electronic fingerprints for both  $s^{\pm}$-wave and $d_{x^2-y^2}$-wave pairing states. We have also studied other pairing symmetries such as $d_{xy}$- and $s_{x^2-y^2}$-wave that are possible for the tetragonal point-group symmetry, and found that they have significantly weaker strength compared to the former two and thus are not further discussed here. Both $s^{\pm}$- and $d_{x^2-y^2}$-wave symmetries are essentially nearest-neighbor pairing but differ by an invariant or broken $C_4$-symmetry, respectively, which governs the basis functions of the SC order parameters $g_{s^{\pm}/d_{x^2-y^2}}=\cos{k_xa}\pm\cos{k_ya}$. The nodal planes of the two pairing states are thus oriented along the $k_x=\mp k_y$-directions as shown in Fig.~\ref{fig1}{g-l}, and they cut through the large squarish FS (band 2) for $s^{\pm}$ pairing, while they intercept with all FSs for $d_{x^2-y^2}$ pairing.

\begin{figure*}[h]
\rotatebox[origin=c]{0}{\includegraphics[width=0.8\textwidth]{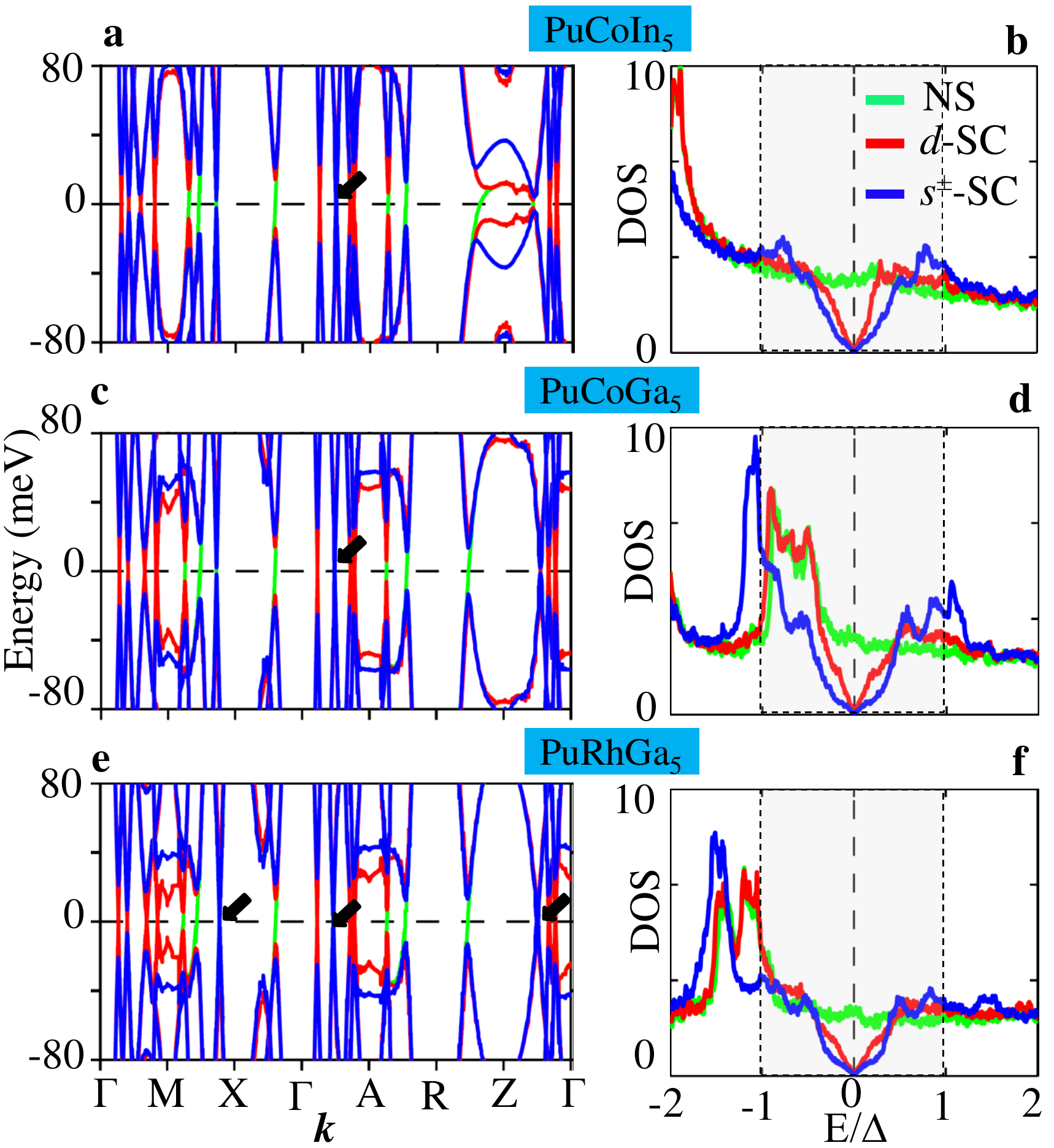}}
\caption{{Gapped electronic DFT dispersions, nodal SC quasiparticle states.}
(a), (c), and (e)
Low-energy electronic dispersions in the normal state with $d$- and $s^{\pm}$-wave gaps plotted along representative high-symmetry directions in the Brillouin zone. Here symbols $\Gamma$ (Z)=($0,0,0/\pi$), M (A) =($\pi,\pi,0/\pi)$, and X (R)=($0,\pi,0/\pi$). Black arrows mark the locations of the zero-gap or nodal lines for the $s^{\pm}$, while those for  $d$ wave are ubiquitous along all $\Gamma-$M, and $\Gamma$-Z directions.
(b), (d), and (f)
Corresponding DOS in the SC state for all three cases discussed in the corresponding left column. Residual nodal states below the SC gap are evident for both pairing symmetries. For ease of comparison an artificial gap amplitude of $\Delta=40$~meV (shaded region) is used in all bands and for all compounds.
} \label{fig3}
\end{figure*}

To be specific, we draw the SC quasiparticle bands $E_{\bf k}^n=[(\xi^n_{\bf k})^2+(\Delta^n_{\bf k})^2]^{1/2}$, where $n$ is the band index, $\Delta^n_{\bf k}=\Delta_0 g({\bf k})$, and $\Delta_0$ is the gap amplitude, 
taken to be the same for all bands and pairing states for direct comparison. Since at the FS each {\bf k} point is uniquely associated with a specific band, $g({\bf k})$ carries an implicit band index $n$ when solving the gap equations. While the gap amplitude is measured to be around 5-10~meV \cite{Curro,PCT}, we used an artificially large value of 40~meV for all systems for better visualization. The results are compared in Fig.~\ref{fig3} (left column) for non-SC band (green), $d$-wave (red) and $s^{\pm}$-wave (blue) pairing along representative high-symmetry momentum directions. We see that gapless quasiparticles evidently occur along the $\Gamma$-M direction for $d$-wave pairing, while a robust node is visible for $s^{\pm}$-wave pairing along the $\Gamma$-A-direction for all three systems (with some accidental nodes along other directions in PuCoGa$_5$). The corresponding density of states (DOS), plotted in Fig.~\ref{fig3} (right column) gives further insight into the energetics of the two pairing states. For the same gap amplitude, we detect that  the $s^{\pm}$-wave pairing has a larger effective gap $\Delta_{\bf k}$, i.e., lower DOS inside the gap, and yields a very much `U'-shaped DOS with nodes at the Fermi level, in contrast to the prototypical `V'-shaped DOS for the $d$-wave pairing. The nodal electronic states of both pairing symmetries are evident in the NMR~\cite{Curro,Sakai}, $\mu$SR~\cite{muSR1,muSR2}, and PCS data~\cite{PCT}, although in these measurements the pairing symmetry has been generally interpreted to be consistent with the single-band $d$-wave pairing symmetry. We anticipate that direct spectroscopies such as angle-resolved photoemission spectroscopy (ARPES), field-angle dependent thermodynamic measurements \cite{Dasfldangle}, and scanning tunneling microscopy and spectroscopy (STM/S) will be able to distinguish between these two pairing symmetries.

\vskip12pt
\noindent
{\bf Normal State Instability and Pairing Strength.} We now carry out calculations of the pairing symmetry and pairing strength based on the spin-fluctuation mechanism of electron pairs in the spin-singlet channel. The methodology of the corresponding calculation is well established for other materials, and we generalize it to be combined with the DFT framework, in which all band structure information such as crystal field splitting, spin-orbit coupling are incorporated in the electronic dispersions. The dominant many-body interactions are onsite Coulomb repulsion for intra-band and inter-band components, which play separate roles for different pairing channels. These are included within the random-phase approximation (RPA). For this building block, all the relevant energetics, coming from the single-particle terms as well as many-body interactions, are incorporated within the multiband anisotropic spin  $\tilde{\chi}^s({\bf k},{\bf k}^{\prime})$ and charge $\tilde{\chi}^c({\bf k},{\bf k}^{\prime})$ susceptibilities defined as $\tilde{\chi}^{s/c}({\bf k},{\bf k}^{\prime})=\tilde{\chi}({\bf k},{\bf k}^{\prime})/(\tilde{1}\mp\tilde{U}^{s/c}\tilde{\chi}({\bf k},{\bf k}^{\prime}))$, where $\tilde{1}$ is the unity matrix, ${\tilde U}^{s/c}$ are the corresponding onsite interaction matrices (defined below), and $\tilde{\chi}({\bf k},{\bf k}^{\prime})$ is the bare interaction defined in Eq.~1 above. All variables with tilde are of matrix dimension $16\times 16$. The corresponding spin-singlet pairing matrix is~\cite{scalapino86,Takimoto}
\begin{equation}
{\tilde \Gamma}({\bf k},{\bf k}^{\prime})= \frac{1}{2}{\rm Re}\big[3{\tilde U}^s{\tilde \chi}^s({\bf k},{\bf k}^{\prime}){\tilde U}^s - {\tilde U}^c{\tilde \chi}^c({\bf k},{\bf k}^{\prime}){\tilde U}^c
+ {\tilde U}^s+{\tilde U}^c\big].
\end{equation}

\begin{figure*}[h]
\centering
\rotatebox[origin=c]{0}{\includegraphics[width=0.9\textwidth]{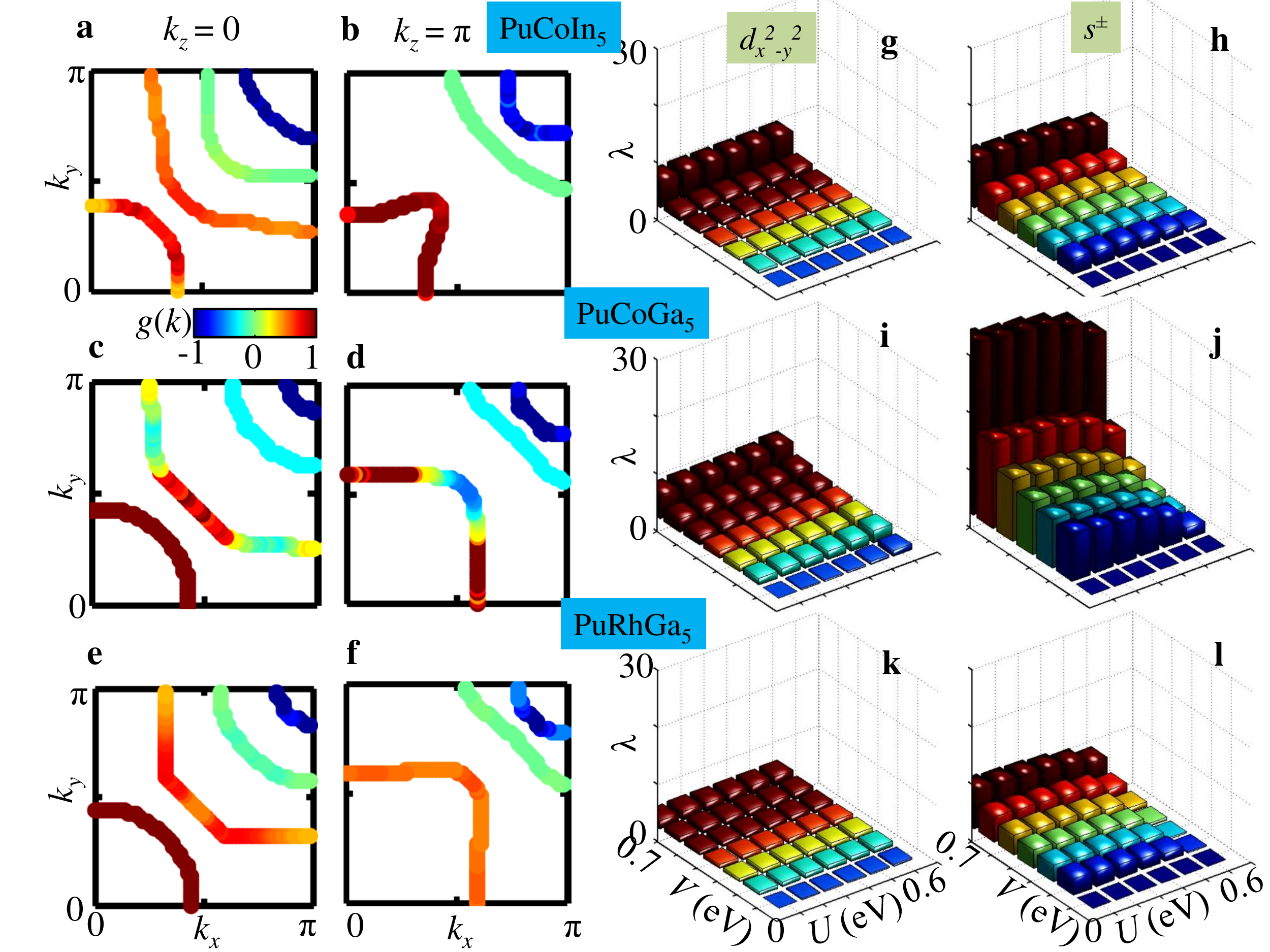}}
\caption{{Pairing eigenfunction and total SC pairing strength $\lambda$.}
(a)-(f), The computed gap function in a color gradient plot overlayed on the corresponding FSs at two representative $k_z$ cuts, computed from Eq.~(\ref{eq:lambda}). (g)-(l) Projected total pairing strength $\lambda=\sum_{nm}\lambda_{nm}$ evaluated from Eq.~(\ref{gammaband}) as a function of $U$ and $V$. The height of squarish bar represents the value of the total pairing strength. Each row describes a different compound, while different columns separate different pairing symmetries. Different colors distinguish same $V$ rows. 
} \label{fig4}
\end{figure*}

Earlier model calculations of this pairing potential in cuprates~\cite{scalapino,Dahm}, heavy-fermion systems~\cite{Takimoto}, organic superconductors~\cite{schmalian}, and pnictides~\cite{Graser}, have produced good estimates of the pairing strength and pairing symmetry, consistent with corresponding experimental data. Following the same strategy, we solve the linearized multiband gap equations by the pairing eigenvalue problem as
\begin{equation}
\lambda g({\bf k}_n)=-\sum_{m,{\bf k}_m^\prime}
\Gamma_{nm}({\bf k}_n,{\bf k}_m^{\prime})g({\bf k}_m^{\prime}),
\label{eq:lambda}
\end{equation}
where ${\bf k}_{n}$ is the momentum for the $n^{\rm th}$ band and so on, and $\Gamma_{nm}$ are the components of the pair vertex in Eq.~(2), after projected into the corresponding band basis.The eigenvalue calculation is performed over the entire three-dimensional FSs to estimate the dominant eigenvalue $\lambda$, and the corresponding eigenvector gives the leading pairing symmetry $g({\bf k})$ (see supplementary materials for the method of calculation). Eq.~(\ref{eq:lambda}) is solved for the representative values of intraband interaction $U$ = 0.5 eV and interband interaction $V$ = 0.5 eV, which yield maximal eigenvalues $\lambda= 2.3, 3.5$ and 2.5 for all three systems in the order discussed, and the corresponding pairing eigenfunctions are plotted in Fig.~\ref{fig4}a-f in a blue to red colormap at the Fermi momenta for two representative $k_z$ cuts (the results are also consistent with other $k_z$ cuts and interaction values). The first point to notice is that there is no clear four-fold symmetry breaking in $g({\bf k})$, which excludes the presence of any significant $d$-wave pairing component. In the supplementary material, we give a detailed fit of the computed $g({\bf k})$ with an $s^{\pm}$ gap function including higher harmonics. Our result indicates that the $k_z$ dependence of $g({\bf k})$ is weak, and also the presence of second and higher harmonics is negligible. Furthermore, we find the expected result that the gap anisotropy is largest in band 2, and then reduces gradually in band 1, 4 and 3. This is expected from the $s^{\pm}$-pairing symmetry as the gap maxima lie at the $\Gamma$ and M points with opposite signs.
 
Next we study the relative strength of various possible pairing channels, and the contributions from each band. We introduce a dimensionless pairing strength by projecting Eq.~(\ref{eq:lambda}) onto a gap function $g_{\alpha}({\bf k})$ with given symmetry $s^\pm$ or d$_{x^2-y^2}$ (denoted by $\alpha$):
\begin{eqnarray}
\lambda_{nm}^\alpha = - \frac{ \oint \frac{d k_n}{v^n_F} \oint \frac{d k_m^{\prime}}{v^m_F} 
g_{\alpha}({\bf k}_n)\Gamma_{nm}({\bf k}_n,{\bf k}_m^{\prime})g_{\alpha}({\bf k}_m^{\prime}) }
{\sum_n (2\pi)^2\oint \frac{d k_n}{v_F^n} g^2_{\alpha}({\bf k}_n)},
\label{gammaband}
\end{eqnarray}
with its total value being $\lambda^{\alpha} = \sum_{n,m} \lambda_{nm}^{\alpha} $. Here $v^n_F$ is the Fermi velocity at momentum $k_n$. The physical interpretation of $\lambda_{nm}$ can be gained qualitatively by studying the peaks in $\Gamma_{nm}({\bf k},{\bf k}^{\prime})$. The dominant pairing potential $\Gamma_{nm}({\bf k},{\bf k}^{\prime})$ is governed mainly by the peaks of $\chi({\bf q}, \omega=0)$, see Fig.~\ref{fig1} at ${\bf q}={\bf k}-{\bf k}^{\prime}$, because they stabilize the gap function that satisfies the condition $g({\bf k})=-g({\bf k}^{\prime})$ at a maximum number of momenta to yield the largest positive pairing strength $\lambda$. Since the dominant peak in $\chi({\bf q},0)$ is determined by the FS scattering enhancement  (or equivalently FS nesting) due to spin fluctuations, the relative strength of pairing symmetry with respect to others depends only on the strength of FS nesting, and is very much insensitive to the specific values of ${\hat U}^{s/c}$. Therefore, we perform the pairing strength calculation for a realistic range of interaction values and draw conclusions based on the robust result from the overall phase diagrams. Based on the prior knowledge from other systems \cite{scalapino,JCDavis}, we know that the intra-band interaction, denoted by $U^2\chi_{nn}$, favors $d$-wave pairing while the inter-band interaction, $V^2\chi_{{nm(\neq n)}}$, enhances the $s^{\pm}$-wave pairing, where $\chi_{{nn}}$ and $\chi_{{nm(\neq n)}}$ are intra- and inter-band susceptibilities with $\chi_{{nm(\neq n)}}>\chi_{{nn}}$ in these compounds. Figure~\ref{fig4}g-l show the total pairing strength $\lambda$ for $d$-wave (left column) and $s^{\pm}$-wave pairing (right column) for three Pu-based superconductors (in three horizontal rows) as a function of $U$ and $V$. In the entire phase diagram, we therefore predict that the $s^{\pm}$-wave pairing dominates over the $d$-wave pairing for all three materials by as large as an order of magnitude. In addition, we also observe that the value of $\lambda$ is maximum in PuCoGa$_5$, which is consistent with its measured value of SC transition temperature $T_c$.

\begin{figure*}[h]
\rotatebox[origin=c]{0}{\includegraphics[width=0.8\textwidth]{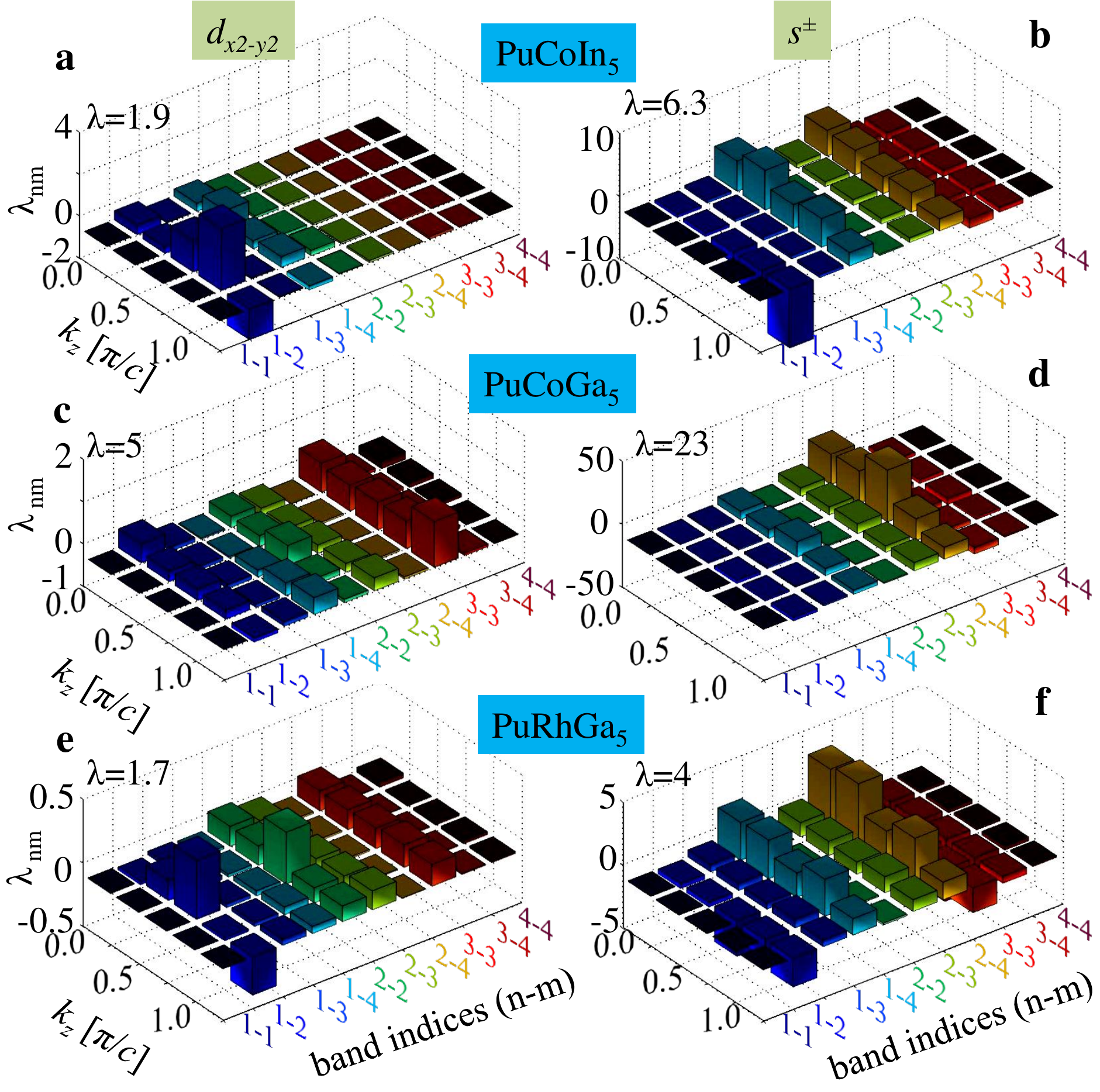}}
\caption{{Band-dependent SC pairing strength for several representative $q_z$ cuts.}
A comprehensive view of band-specific pairing eigenvalues $\lambda_{nm}$ for fixed Coulomb interactions $U=V$=0.5~eV in all bands. In agreement with FS topologies and hot-spots, the inter-band pairing between bands 1,2 to 3,4 contributes a large value to the $s^{\pm}$-wave pairing. For the $d_{x^2-y^2}$-wave pairing, all bands contribute finite values with stronger contributions coming from $\lambda_{22}$ and $\lambda_{33}$, etc. }
\label{fig5}
\end{figure*}

To further delineate the reasons for having a strong $s^{\pm}$-wave pairing channel, we investigate the contributions of each band and the hot-spot wavevector in Fig.~\ref{fig5}. We recall the relevant facts pertaining to the FS topology discussed in Fig.~\ref{fig1} that the dominant FS instability, which also induces the leading pairing instability, commences between bands 1 and 2 to bands 3 and 4 [connected by wavevector $Q\sim(\pi,\pi,q_z)$]. The sign reversal of the pairing states, which is essential for yielding positive $\lambda$, occurs only in the inter-band components between bands 1,2 and bands 3,4 for the $s^{\pm}$-wave pairing symmetry (not for intra-band except band 2 or between bands 1 and 2, or 3 and 4), while for the $d$-wave pairing symmetry both intra- and inter-band components contribute. These facts manifest themselves in the value of $\lambda_{nm} (q_z)$, plotted in Fig.~\ref{fig5}. For the $s^{\pm}$-wave pairing, finite positive pairing strength thus occurs for $\lambda_{13}$, $\lambda_{23}$, $\lambda_{14}$, and $\lambda_{24}$, while the others contribute negative or small pairing strength. Because these pairing components are supported by the strong FS scattering enhancements, the total pairing strength $\lambda$ eventually gains a positive and large value. This scenario should be contrasted with the $d$-wave pairing, in which the total $\lambda$ is positive but smaller in amplitude lest all components of $\lambda_{nm}$ (except $\lambda_{12}$ in the $q_z\sim\pi$-plane) contributing positive values. This is due to the fact that  the $d$-wave pairing, breaking $C_4$ symmetry, obtains sign reversal in each quadrant of the Brillouin zone, thus the momentum sum in Eq.~(3) amounts to a lower amplitude in $\lambda$. On the other hand, the $s^{\pm}$-wave pairing channel causes fairly isotropic and single-sign pairing in each band (except in band 2, which cuts through the nodal plane), and thus contributes large phase space of positive values in the momentum sum. These are the key reasons why the nodal $s^{\pm}$-wave pairing is favored over the $d$-wave pairing channel, although both pairing symmetries attain positive values, and are possible contenders for unconventional superconductivity in the Pu-115 family. As mentioned earlier, the $d_{xy}$- or $s_{x^2-y^2}$-wave pairing symmetries also obtain positive $\lambda$, but much weaker in strength and thus are losing contenders in these systems.

\vskip12pt
\noindent
{\bf Magnetic Resonance Mode and Signatures of Pairing Symmetry.} We present experimentally verifiable signatures for both dominant pairing symmetries. In this context, the magnetic resonance mode is widely considered to be a deciding feature for unconventional pairing symmetry. A well defined spin resonance is observed in cuprates~\cite{Bourges}, iron pnictides~\cite{Christianson}, and Ce-based heavy fermions\cite{Stock}, which is located at characteristic energy and momentum in the SC state. It was shown by Yu {\it et al.}~\cite{Greven} that the resonance energy scales almost linearly with the SC gap in all these materials, suggesting further that the magnetic resonance mode is indeed a feedback effect of the unconventional gap symmetry. Motivated by this universal scaling, we study the evolution of the magnetic excitation spectrum of both SC states and evaluate their characteristics to guide experimental detection.

\begin{figure*}[h]
\rotatebox[origin=c]{0}{\includegraphics[width=0.8\textwidth]{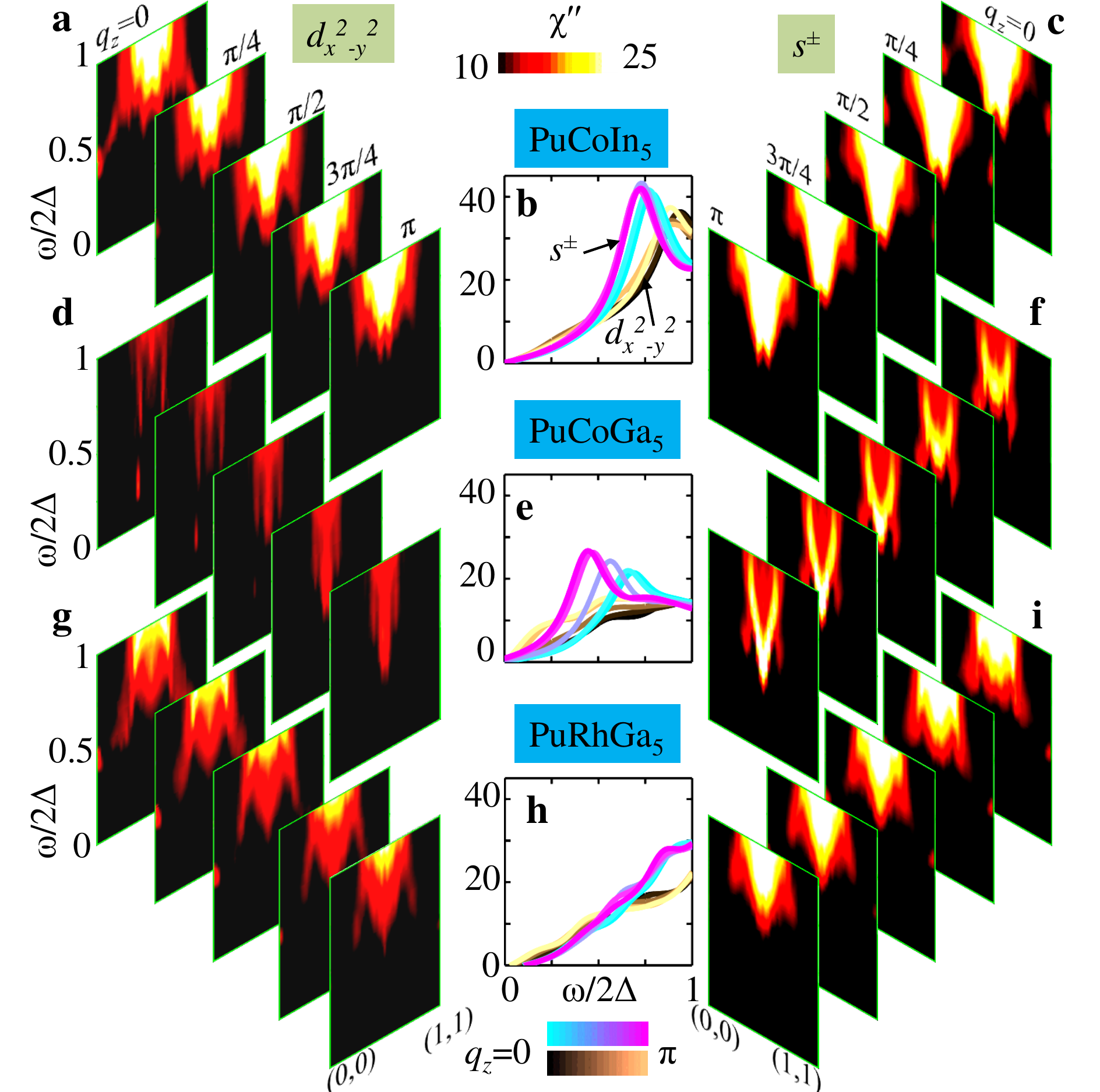}}
\caption{{Evolution of computed magnetic excitation spectra for both pairing symmetries (left and right columns) and all three compounds (rows).}
Each slice of the colormap images shows the imaginary part of the BCS susceptibility within the RPA method, ${\rm Im}\chi_{\text{\tiny BCS-RPA}}$, along  wavevector ${\bf q}= (0,0,q_z)\rightarrow(\pi,\pi,q_z)$, with different slices for different $q_z$ values. The middle panel plots a single cut ${\bf Q}=(\pi,\pi,q_z)$ as a function of excitation energy $\omega$ to visualize the feature of the resonance mode at an energy scale $\omega<2\Delta$, where $\Delta$ is the SC gap amplitude. We set $U=V=$0.4~eV for the resonance calculation.
} \label{fig6}
\end{figure*}

The magnetic resonance spectrum in the SC state of  single-band and multiband systems is well studied within the BCS theory~\cite{scalapino86,schmalian,Dahm,Takimoto,Graser}. A generalization to the DFT band structure is obtained here (see supplementary material for details). The magnetic collective mode is a manifestation of many-body interactions, which is captured within the BCS-RPA framework. In this framework, the RPA formulas remain the same as before, while the bare susceptibility is replaced by the BCS susceptibility which involves additional terms coming from particle-particle, and hole-hole scattering process. The magnetic resonance calculation uses the full BCS-RPA susceptibility as shown in Fig.~\ref{fig6}. However, to obtain a qualitative understanding of the fundamental energy and momentum scale of the resonance, one can use a simplified expression to estimate the resonance condition $\omega_{nm}^{res}({\bf Q})=|\Delta^n_{{\bf k}_F}|+|\Delta^m_{{\bf k}_F+{\bf Q}}|$, given that ${\rm sgn}[\Delta^n_{{\bf k}_F}] = - {\rm sgn}[\Delta^m_{{\bf k}_F+{\bf Q}}]$, where ${\bf k}_F$ are those Fermi hot-spots, which provide strong nesting for wavevector ${\bf Q}$. Such an analysis has been successfully used before for cuprates~\cite{Dascuprate} and pnictides~\cite{Daspnictide}, with its quantitative value and intensity subject to the details of the band structure and the orbital overlap of matrix-element parameters.

Clearly, the condition for having a strong resonance mode has the same underlying mechanism as that of the positive pairing strength discussed earlier. Consistent with the aforementioned discussion, we thus expect to have spin resonance in the vicinity of ${\bf Q}=(\pi,\pi,q_z)$, which involves a sign reversal in both $d$-wave and $s^{\pm}$-wave pairing symmetries. Our results of magnetic resonance spectra are shown in Fig.~\ref{fig6} for both pairing symmetries. As for the value of $\lambda$ in both cases, the intensity of the magnetic excitation spectrum is weaker and more spread out over the momentum space for the $d$-wave pairing case, while it is substantially more localized around ${\bf Q}$ with maximum intensity shifted towards $q_z\rightarrow\pi/c$ for the $s^{\pm}$-wave pairing case. To affirm our statement, in Fig.~\ref{fig6} we plot the total $\chi({\bf Q},\omega)$ (the energy axis is normalized by the SC gap amplitude to perform a comparative study between different materials with different $T_c$) along the diagonal direction and at five representative $q_z$ cuts. Also the single-momentum cuts at $(\pi,\pi,q_z)$ are plotted for both pairing symmetries with different colormaps distinguishing different $q_z$ values in the middle column, while different rows are for different materials. We clearly see the so-called localized collective mode  for the $s^{\pm}$-wave pairing in both PuCoIn$_5$, PuCoGa$_5$, but not in PuRhGa$_5$ system. Our prediction of the ratio $\omega_{res}/2\Delta\sim$0.5-0.75 is in reasonable agreement with universal scaling\cite{Greven} and can be verified by inelastic neutron scattering measurements, which has the ability to detect both the energy and momentum resolved collective ${\bf S}=1$ spin excitations.
\\
\vskip10pt
\noindent
{{\bf PCS Results.}} To elaborate on the spectroscopic fingerprints of both pairing symmetries, we calculate their respective PCS spectra using the Blonder-Tinkham-Klapwijk (BTK) formalism~\cite{BTK}  generalized to multiband systems with anisotropic FSs and SC order parameters~\cite{PCSBrinkman,PCSDaghero}, see supplementary materials. In order to keep the problem tractable, we consider only normal incidence of electrons from the metallic tip and neglect interband transitions as well as Fermi velocity mismatch between the tip and the Pu-115 compounds. The corresponding results are given in Fig.~\ref{fig7} and compared with available conductance data for PuCoGa$_5$~\cite{PCT} in Fig.~\ref{fig7}{b}. We see that for both nodal $d$-wave and $s^{\pm}$-wave pairing symmetries, the calculated conductance spectrum exhibits a characteristic zero-bias conductance peak, which marks the presence of nodes and the hallmark of Andreev bound states. To contrast these results, we also calculate the PCS spectrum for isotropic (nodeless) $s$-wave pairing (green line) which shows a suppressed Andreev reflection signal for finite interface barrier potential. For a reasonable parameter choice of $\Delta_0$=10 meV, interface transparency coefficient $Z$=1.55, and a rotation angle $\alpha=\pi/8$ (of the crystallographic $a$ axis with respect to the normal of the interface), we can fit the experimental data very well with nodal $s^{\pm}$-wave pairing symmetry. Of course, it is not impossible to fit the data with $d$-wave with another parameter choice even in this realistic multiband model. This implies that PCS conductance data are consistent with nodal gap functions, but cannot unequivocally determine the locations of nodes.

\vskip12pt
\noindent
{\bf Outlook.} Obtaining a consistent theory of unconventional superconductivity, which can describe cuprate, pnictide, organic, heavy-fermion, as well as actinide superconductors has a pressing need. Considerable
consistency is achieved so far in all three former families of superconductors~\cite{scalapino,JCDavis,scalapino86,Takimoto,Dahm,schmalian,Graser} in terms of spin-fluctuation-mediated superconductivity, pairing symmetry, and magnetic resonance mode, which motivated us to perform these studies in the actinide family. Here we provided the first DFT-based spin-fluctuation calculation of the pair symmetry in the three-dimensional, multiband actinide superconductor family and find the surprising result of the dominant $s^{\pm}$-wave pairing symmetry with a nodal gap, and not the so often assumed $d$-wave gap. The feedback effect of this unconventional pairing yields a strong magnetic resonance mode, which can be tested in future inelastic neutron scattering measurements. In the past, the $d$-wave gap was mostly proposed because it was the simplest scenario based on a single band that could explain power laws in the low-temperature behavior of specific heat, spin-lattice-relaxation rate, and magnetic penetration depth. Of course, gap nodes on the FS have a profound influence on electronic excitations and the formation of Andreev bound states, which provide a natural explanation of the observed zero-bias conductance peak in the point-contact spectra \cite{PCSDaghero,PCSGoll,PCSYu}. Interestingly, the observed zero-bias conductance peak can be fit equally well with the nodal $s^{\pm}$-wave pairing symmetry as shown here. The identification of the nodal $s^{\pm}$-wave pairing symmetry will also provide insight into the physics of iron-pnictide superconductors, which are believed to host unconventional pairing symmetry. Therefore, we envisage that further studies of this actinide family of unconventional multiorbital superconductors will advance the lofty goal of obtaining a unified spin-fluctuation picture of superconductivity.

\begin{acknowledgments}
This work was supported by the U.S.\ DOE under Contract No.\ DE-AC52-06NA25396 at the Los Alamos National Laboratory through the LDRD Program and completed with support by the Office of Basic Energy Sciences, Division of Materials Sciences and Engineering. We are grateful for the computational resources of the National Energy Research Scientific Computing Center (NERSC), which is supported by the Office of Science of the U.S.\ DOE under Contract No.\ DE-AC02-05CH11231.
\end{acknowledgments}

\noindent
\newpage

\vskip12pt
\noindent
{\bf Author Contribution.}
T.D. has provided the idea of the research, performed the calculations, and analysis and written the paper. J.X.Z. has performed the first-principles band structure calculation and contributed to the discussions. M.J.G. has contributed to the calculation of Point contact spectroscopy, and in the analysis and in the writing. All authors have revised the manuscript.  

\vskip12pt
\noindent
{\bf Competing Financial Interest.}
The authors declare no competing financial interest in this manuscript.

\noindent
\newpage

%

\noindent
\newpage

\noindent
\newpage

%

\noindent
\newpage

%

\noindent
\newpage

%

\noindent
\newpage

\newpage

\begin{figure}[h]
\rotatebox[origin=c]{0}{\includegraphics[width=0.45\textwidth]{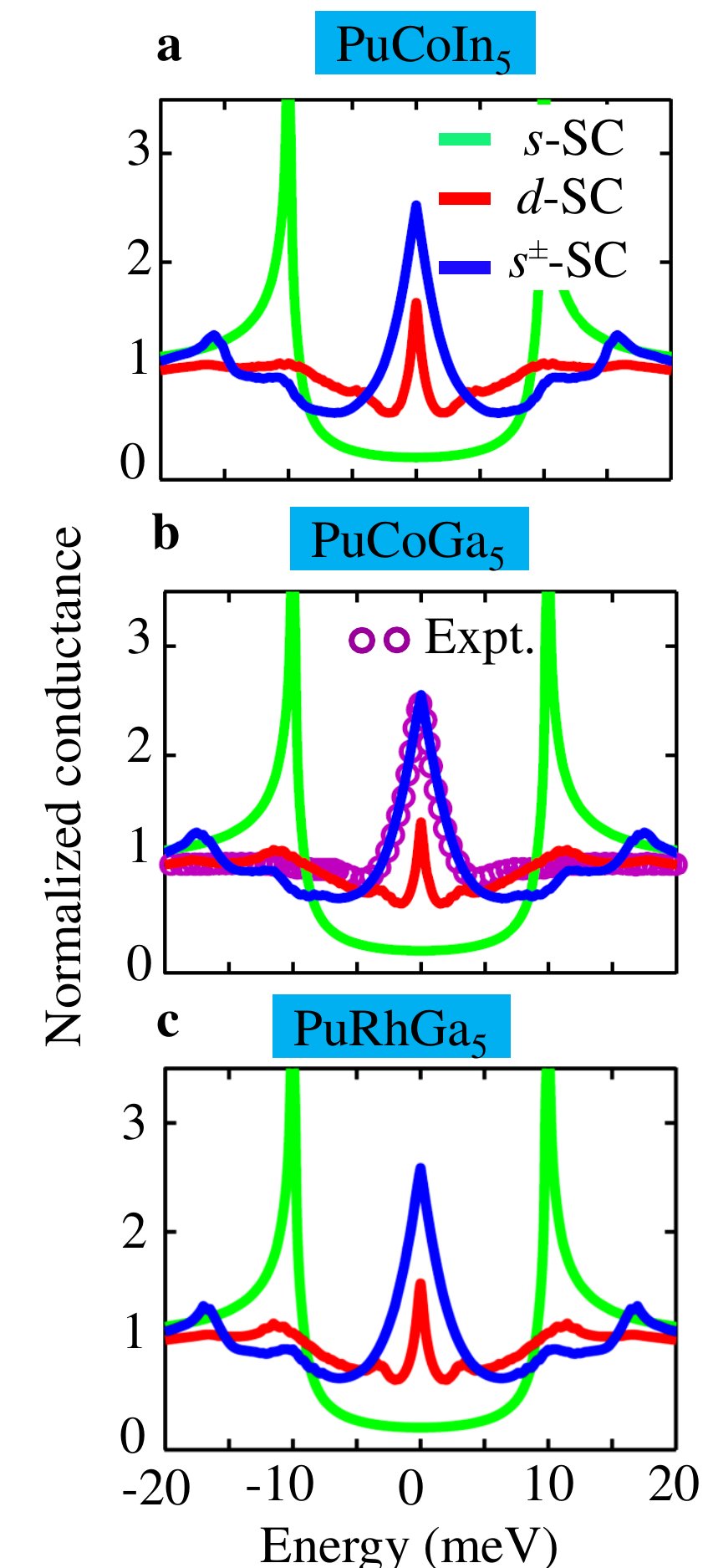}}
\caption{PCS conductances.
Computed PCS spectra for nodal $d_{x^2-y^2}$ (red), $s^\pm$ (blue), and  isotropic $s$ (green) gaps using a generalized multiband BTK formalism.
All data are normalized to their normal-state conductance. A zero-bias conductance peak is seen in both nodal $d_{x^2-y^2}$- and  $s^{\pm}$-wave gaps, but not in the fully gapped isotropic $s$-wave gap. 
}\label{fig7}
\end{figure}

\clearpage
\newpage

{\Large{Supplementary Materials}}

\section{I. Density functional theory calculations.}
We have performed electronic structure calculations of these materials within the framework of DFT and the results are shown in Fig.~1 of the main text. Our calculations were carried out by using the full-potential linearized augmented plane wave (FP-LAPW) method as implemented in the WIEN2k code~\cite{PBlaha2001}. The generalized gradient approximation (GGA)~\cite{JPPerdew1996} was used for the exchange-correlation functional. The spin-orbit coupling was included in a second variational way, for which relativistic $p_{1/2}$ local orbitals were added into the basis set for the description of the $6p$ states of plutonium \cite{Kunes2001}. The muffin-tin radii were: 2.5$a_0$ for Pu, 2.46$a_0$ (PuCoGa$_5$) and  2.5$a_0$ (PuCoIn$_5$) for Co, 2.5$a_0$ for Rh, 2.18$a_0$ (PuCoGa$_5$) and 2.23$a_0$ (PuRhGa$_5$) for Ga, and 2.39$a_0$ for In, where $a_0$ is the Bohr radius. The energy spread to separate the localized valence states was -6 Ryd. The criterion for the number of plane waves was $R_{MT}^{{\rm min}} K^{{\rm max}} = 8$ and the number of $\mathbf{k}$-points was $40\times 40 \times 25$. The experimentally determined crystallographic structures were used~\cite{PuCoGa5,PuRhGa5,PuCoIn5}.

\section{II. Point-contact spectroscopy calculation using anisotropic Fermi surfaces and anisotropic order parameter}



We calculate the point-contact spectrum (PCS) by taking into account the full Fermi surface (FS) anisotropy of a multiband system with anisotropic order parameter by following the formalism given in Refs.~\cite{Kashiwaya,PCSMazin,PCSBrinkman,PCSDaghero}. For simplification, however, we only include the anisotropy in the FS of the superconducting (SC) material, while that of the normal metal tip is neglected. Let us define ${\bf n}$ as the unit vector in the direction of the total injected current, which for simplicity we choose to be perpendicular to the contact interface between the SC (S) and normal metal (N) interface. As a consequence the components  along the direction ${\bf n}$ of
the Fermi velocities at wavevector ${\bf k}$ in the $i^{\rm th}$ FS sheet of the superconductor are ${\bf v}_{i{\bf k}}\cdot {\bf n} = v_{i{\bf k},n}$, where
${\bf v}_{i{\bf k}}=-\frac{1}{\hbar} \nabla_{\bf k} E_{i{\bf k}}$, and $E_{i{\bf k}}$ is the corresponding quasiparticle band. Generalizing the Blonder, Tinkham and Klapwijk (BTK) formula \cite{BTK} to anisotropic FSs of multiband superconductors, it was shown that the total normalized conductance seen along the direction ${\bf n}$ can be written as \cite{PCSDaghero}
\begin{eqnarray}
\left\langle G(E)\right\rangle_{I\parallel{\bf n}}=\frac{\sum_i\left\langle \sigma_{i{\bf k}} (E) D_{i{\bf k}} v_{i{\bf k},n}\right\rangle_{{\rm FS}_i}}{\sum_i\left\langle D_{i{\bf k}} v_{i{\bf k},n}\right\rangle_{{\rm FS}_i}} ,
\label{totconductance}
\end{eqnarray}
where $D_{i{\bf k}}=1/v_{i{\bf k}}$ is the density of states on the $i^{\rm th}$ FS sheet, and $\sigma_{i{\bf k}} (E)$ is the BTK SC transition probability calculated as follows. Here we neglect interband interference effects and variations in the tunneling matrix elements (different weight factors) between the normal tip and different bands of the superconductor.
\begin{figure}
\rotatebox[origin=c]{0}{\includegraphics[width=.95\columnwidth]{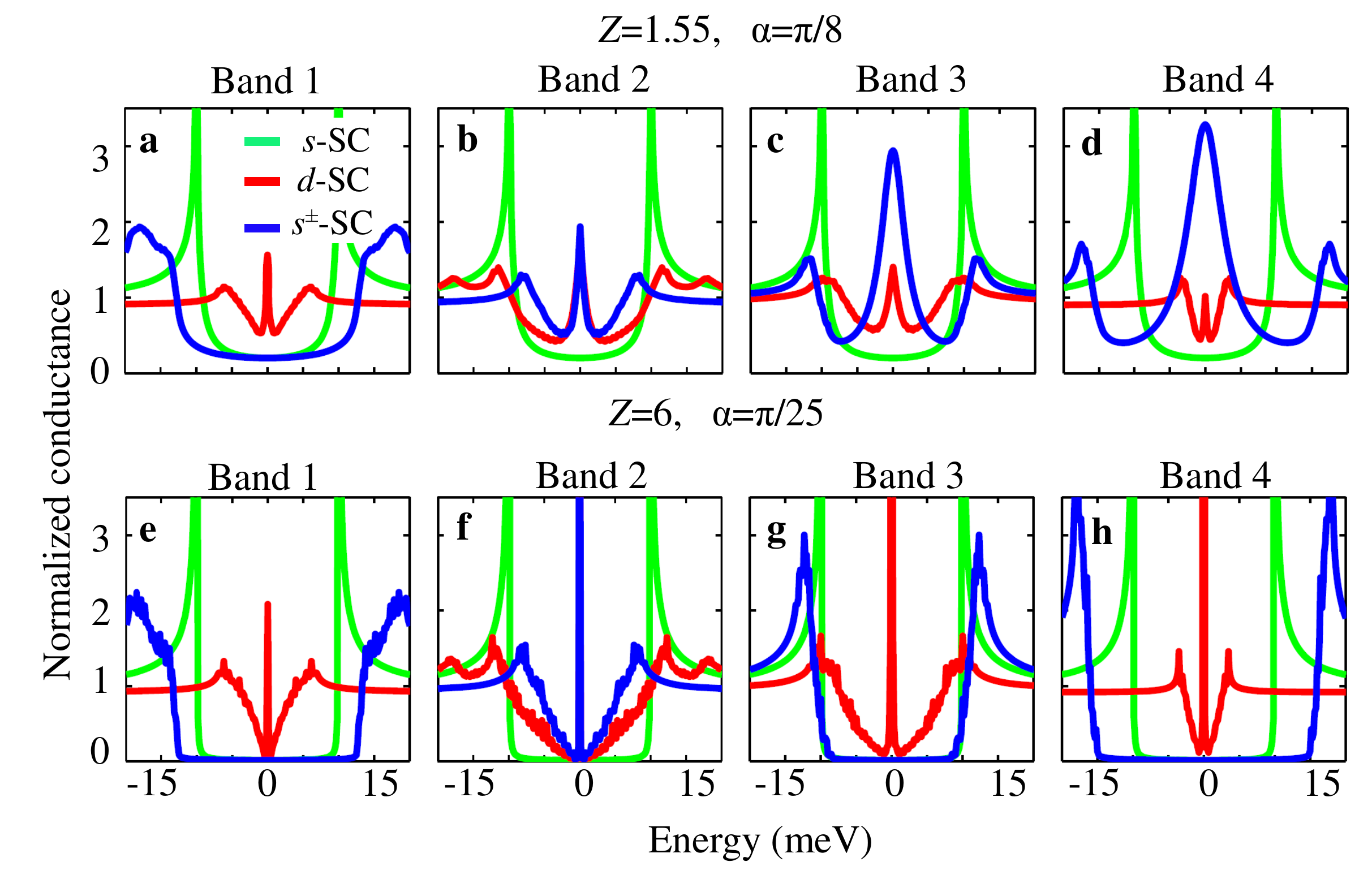}}
\vspace{16pt}
\parbox{\textwidth}{
{\bf Fig.~S1}: Calculated PCS spectra, decomposed in different bands for two parameter sets for PuCoGa$_5$. The zero-bias conductance peak is present in all bands for $d$-wave pairing and in band 2 for $s^{\pm}$-pairing as expected. For a larger value of $\alpha$, we find that a zero-bias conductance peak is induced in bands 3 and 4 in $s^{\pm}$-wave pairing, in addition to band 2. Since band 3 and 4 have larger gap amplitudes which enables a broader PCS conductance peak and match the experimental data, as shown in (a)-(d). For $\alpha=\pi/25$, the conductance peak only survives in band 2 for this pairing.}
\label{fig7}
\end{figure}

Let us assume that $\theta_{i{\bf k}}$ is the transmission angle at the interface between the normal and SC materials for band $i$ at Fermi momentum ${\bf k}$. The specific SC gap function is defined in terms of $\theta_{i{\bf k}}$ as $\Delta_{i{\bf k}}(\theta_{i{\bf k}})=\Delta_{0}(\cos{k_x^i}\pm\cos{k_y^i})$, for $s^{\pm}$- and $d$-wave pairing, respectively and $k_{x,y}^i$ are the Fermi momentum for the $i^{th}$-band. Then $\theta_{i{\bf k}}={\rm tan}^{-1}(k_y^i/k_x^i)$. Lets us also define $\alpha$ as the rotation of the crystallographic $a$-axis with respect to the normal to the interface ($x$ axis). In this circumstance, the electron-like and hole-like quasiparticle (EQs/HQs) injected in the SC material with angles $\pm\theta_{i{\bf k}}$, they access different gap values as $\Delta^{\pm}_{i{\bf k}}=\Delta_{i{\bf k}}(\pm\theta_{i{\bf k}}-\alpha)$. In this case the SC transition probability becomes
\begin{eqnarray}
\sigma_{j{\bf k}} (E) = \tau_{\rm N} \frac{1+\tau_{\rm N} \mid \gamma^+_{j{\bf k}}(E)\mid^2 + (1-\tau_{\rm N}) \mid\gamma^+_{j{\bf k}}(E)\gamma^-_{j{\bf k}}(E)\mid^2}{\mid 1+(1-\tau_{\rm N}) \mid\gamma^+_{j{\bf k}}(E)\gamma^-_{j{\bf k}}(E)\exp{(i\phi_{i{\bf k}})}\mid^2},
\end{eqnarray}
where the function
\begin{equation}
\gamma^{\pm}_{i{\bf k}}(E)=\frac{\mid E\mid-\sqrt{E^2-\mid \Delta^{\pm}_{i{\bf k}}\mid^2 }}{\mid \Delta^{\pm}_{i{\bf k}}\mid},
\end{equation}
and $\phi_{i{\bf k}}=\phi^-_{i{\bf k}}-\phi^+_{i{\bf k}}$, with $\phi^{\pm}_{i{\bf k}}$ being the phases of $\Delta^{\pm}_{i{\bf k}}$. It should be noted that $\gamma^{\pm}_{i{\bf k}}(E)$ are, in general, complex functions even if the momentum-dependent gap $\Delta_{i{\bf k}}$ is real when $E<\Delta_{i{\bf k}}$. For simplicity, we further assume a band-independent and momentum-independent interface barrier, the parameter $\tau_{\rm N}$ is the transparency factor of the barrier in the BTK approximation of current injection perpendicular to the SN interface, defined as $\tau_{\rm N}=1/(1+Z^2)$. The limit $Z=0$ gives the perfectly transparent junction in the transmission limit, i.e., the ideal Andreev reflection regime.

\begin{figure}
\rotatebox[origin=c]{0}{\includegraphics[width=.8\columnwidth]{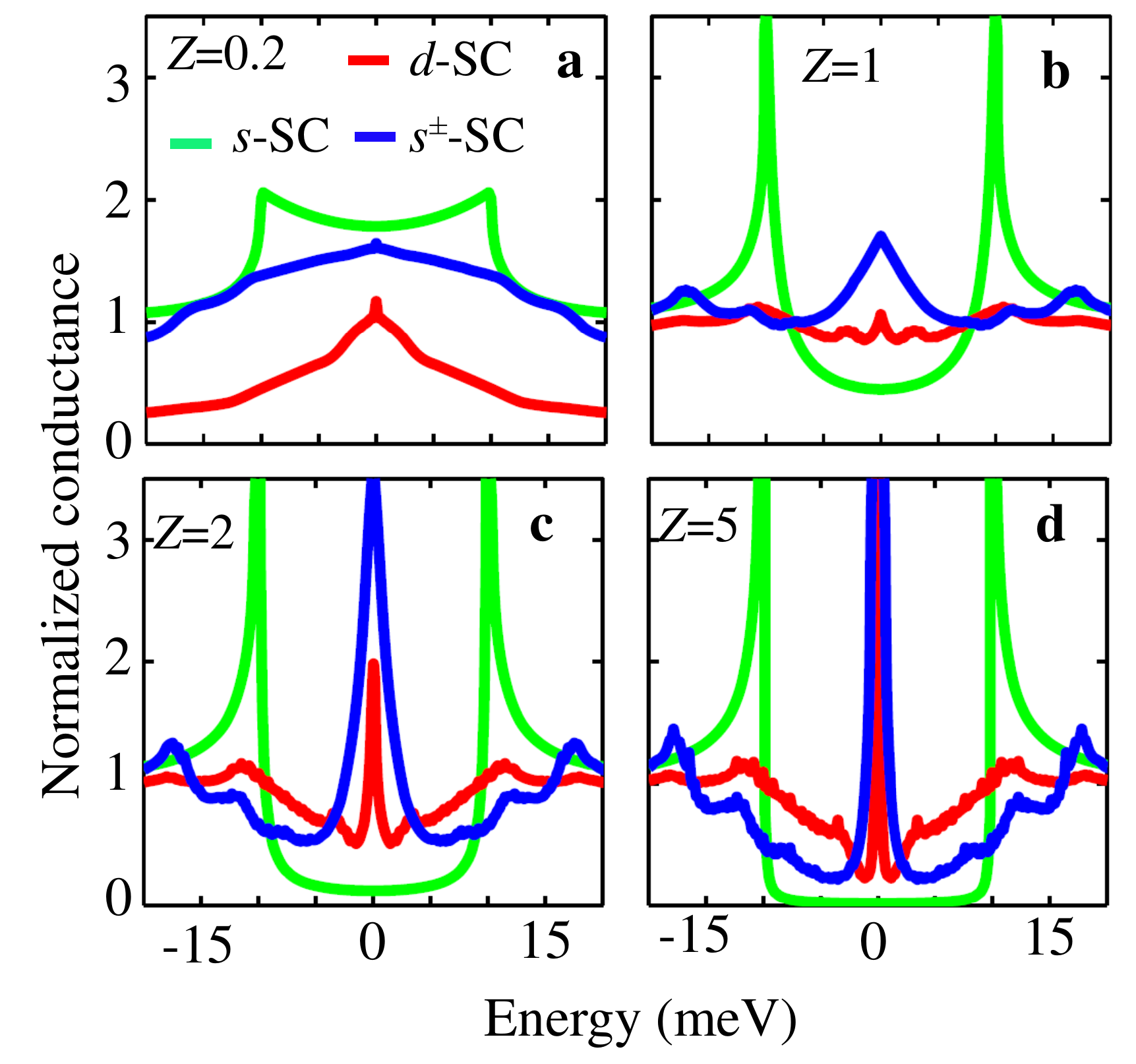}}
\vspace{16pt}
\parbox{\textwidth}{
{\bf Fig.~S2}: Calculated PCS spectra at different parameters of the interface barrier potential $Z$ with $\alpha=\pi/8$, and $\Delta_0$=10~meV.  }
\label{fig9}
\end{figure}

In Fig.~6 of the main text, we presented results of the PCS conductance for the nodal $d_{x^2-y^2}$-wave, nodal $s^{\pm}$-wave pairing symmetries, and compared these results with a nodeless and fully isotropic $s$-wave pairing symmetry. For the best fit to the available experimental data of PuCoGa$_5$,\cite{Daghero}, we set $Z=1.55$, $\alpha=\pi/8$ and a fixed SC gap amplitude of $\Delta_0=10$~meV for all three pairing symmetries. We showed that a single and sharp zero-bias peak is reproduced for $s^{\pm}$-pairing. Of course, even in this multiband setup, it may be possible to fit  the experimental data with $d$-wave pairing.

In Fig.~S1, we showed the band decomposed PCS spectra for two different parameter sets for three different parameter sets. The evolution of the PCS spectrum for different values of $Z$ is discussed in Fig.~S2.

\section{III. Multiband coulomb interactions}
For intermetallic actinides, the spin-orbit coupling is very strong and causes a band splitting of the 5$f$ states of about 1~eV.
This quenches the Hund's coupling term, because $J_H\ll \lambda_{SOC}$. The remaining interaction terms account for the onsite intra- and inter-band Coulomb repulsions as given by the interaction Hamiltonian
\begin{eqnarray}
H_{int} = \sum_{{\bf k},{\bf k}^{\prime}}\left[\sum_{n} U_{n} c^{n\dag}_{{\bf k}\uparrow}c^n_{{\bf k}\uparrow}c^{n\dag}_{{\bf k}^{\prime}\downarrow}c^n_{{\bf k}^{\prime}\downarrow}
+\sum_{n\ne m,\sigma,\sigma^{\prime}} V_{nm} c^{n\dag}_{{\bf k}\sigma}c^n_{{\bf k}\sigma}c^{m\dag}_{{\bf k}^{\prime}\sigma^{\prime}}c^m_{{\bf k}^{\prime}\sigma^{\prime}}\right].
\end{eqnarray}
Here $c^{n\dag}_{{\bf k}\sigma} (c^{n}_{{\bf k}\sigma})$ creates (annihilates) a Bloch state at momentum ${\bf k}$, (pseudo-) spin $\sigma=\uparrow/\downarrow$ in the $n^{\rm th}$-band. The interaction matrices $\tilde{U}^{s/c}$ used in the RPA formalism consist of components $U_n$ and $V_{nm}$. The different bandwidths of different bands amount to different critical values of $U$ and $V$, determined by the positive RPA denominator as $\tilde{U}^{s/c}_{nm}\tilde{\chi}_{nm}\le 1$. Using this condition we find that the critical interaction values for bands 2 and 3 are considerably small and we fix these values to be $2U_{2}=V_{13}=V_{23}=200$ meV at all points for the calculations of $\lambda$ in Fig.~3. The rest of the interactions are taken to be same for all bands as $U_n=U$, and $V_{nm}=V$. In drawing the parameter space in Fig.~3, we limit the values of $U$ and $V$ to below 600 meV and 700 meV, respectively, because for larger values the RPA susceptibilities for bands 3 and 4 become negative, which is above the critical values set by the RPA denominator. However the general conclusion of the dominant $s^{\pm}$-wave pairing than $d$-wave pairing symmetry strength $\lambda$ is consistent throughout the $U-V$ map. Since the essential physics is determined by the FS topology and nesting conditions, we anticipate that this conclusion also remains valid for higher values of $U$ and $V$.

\section{IV. BCS susceptibility and spin resonance}

In this section, we give the details of our spin susceptibility calculation presented in the main text. All ``super" matrices with tilde are defined as ${\tilde \chi}_{ij}=\chi_{nm}\delta_{ij}$, etc., with super indices $i=4(n-1)+m$,  and $n, m=1-4$ are band indices. Thus variables with tilde have matrix dimension $16\times 16$. We follow closely earlier work \cite{Takimoto,Graser} and evaluate the spin-resonance susceptibility in the SC state within the random phase approximation (RPA) of the BCS formalism, which is given by the bare bubble transverse spin susceptibility
\begin{widetext}
\begin{eqnarray}
\chi_{nm}({\bf q},\omega) &=& \int \frac{d {\bf k}}{\Omega_{\rm BZ}}~M_{nm}({\bf k},{\bf q}) \left\{\frac{1}{2}\left[1+\frac{\xi^{n}_{\bf k}\xi^{m}_{{\bf k}+{\bf q}} + \Delta^{n}_{\bf k}\Delta^{m}_{{\bf k}+{\bf q}} }{E^{n}_{\bf k}E^{m}_{{\bf k}+{\bf q}}}\right]\frac{f(E^{n}_{\bf k})-f(E^{m}_{{\bf k}+{\bf q}})}{\omega-E^{n}_{\bf k}+E^{m}_{{\bf k}+{\bf q}}+i\delta}\right.\nonumber\\
&& \hskip0.cm + \frac{1}{4}\left[1+\frac{\xi^{n}_{\bf k}}{E^{n}_{\bf k}}-\frac{\xi^{m}_{{\bf k}+{\bf q}}}{E^{m}_{{\bf k}+{\bf q}}}     - \frac{\xi^{n}_{\bf k}\xi^{m}_{{\bf k}+{\bf q}} + \Delta^{n}_{\bf k}\Delta^{m}_{{\bf k}+{\bf q}} }{E^{n}_{\bf k}E^{m}_{{\bf k}+{\bf q}}}\right]\frac{1-f(E^{n}_{\bf k})-f(E^{m}_{{\bf k}+{\bf q}})}{\omega+E^{n}_{\bf k}+E^{m}_{{\bf k}+{\bf q}}+i\delta} \nonumber\\
&&\hskip0.cm \left.+ \frac{1}{4}\left[1 - \frac{\xi^{n}_{\bf k}}{E^{n}_{\bf k}}+\frac{\xi^{m}_{{\bf k}+{\bf q}}}{E^{m}_{{\bf k}+{\bf q}}} -\frac{\xi^{n}_{\bf k}\xi^{m}_{{\bf k}+{\bf q}} + \Delta^{n}_{\bf k}\Delta^{m}_{{\bf k}+{\bf q}} }{E^{n}_{\bf k}E^{m}_{{\bf k}+{\bf q}}}\right]\frac{f(E^{n}_{\bf k})+f(E^{m}_{{\bf k}+{\bf q}})-1}{\omega-E^{n}_{\bf k}-E^{m}_{{\bf k}+{\bf q}}+i\delta} \right\}
\nonumber\\
\label{seq:1}
\end{eqnarray}
\end{widetext}
\noindent
and the RPA susceptibility is attained with the onsite Coulomb interaction super matrix, $\tilde U$,
\begin{eqnarray}
\tilde\chi_{RPA}({\bf q},\omega) &=&  \left[ 1 - \tilde U \tilde\chi({\bf q},\omega) \right]^{-1} \tilde\chi({\bf q},\omega) .
\label{seq:2}
\end{eqnarray}
Here $E^{n}_{\bf k}=[(\xi_{\bf k}^{n})^2+(\Delta_{\bf k}^{n})^2]^{1/2}$ is the SC quasiparticle energy of the eigenstate $\xi_{\bf k}^{n}$ ($n$ is the band index) and $\Delta_{\bf k}^{n}$ is the SC gap function. $M_{nm}({\bf k},{\bf q})$ is the matrix element consisting of the eigenstates of the initial and final scattered quasiparticle states. In Eq.~\ref{seq:1}, the first term is called particle-hole scattering term, which vanishes for $\omega\le 2\Delta$, regardless of the pairing symmetry due to particle-hole symmetry. The second and third terms are for particle-particle and hole-hole scattering, respectively, which become active in the SC state. Focusing on the third term (a similar analysis applies to the second term), we find that this term contributes a non-zero value only when
sign$[\Delta^{n}_{\bf k}]\ne$sign$[\Delta^{m}_{{\bf k}+{\bf q}}]$
(since $\xi^{n}_{\bf k} =0$ on the Fermi surface). A pole is thus obtained in the imaginary part of $\chi_{nm}$ at
\begin{eqnarray}
\omega^{\rm res}_{nm}({\bm q}) = |\Delta^{n}_{\bf k}|+|\Delta^{m}_{{\bf k}+{\bf q}}|.
\end{eqnarray}
Of course, the many-body  and  matrix-element effects can shift the energy scale as discussed in the Method section in the main text.

\section{III. Computed gap function and weak higher-order harmonics}
\begin{figure}
\rotatebox[origin=c]{0}{\includegraphics[width=.95\columnwidth]{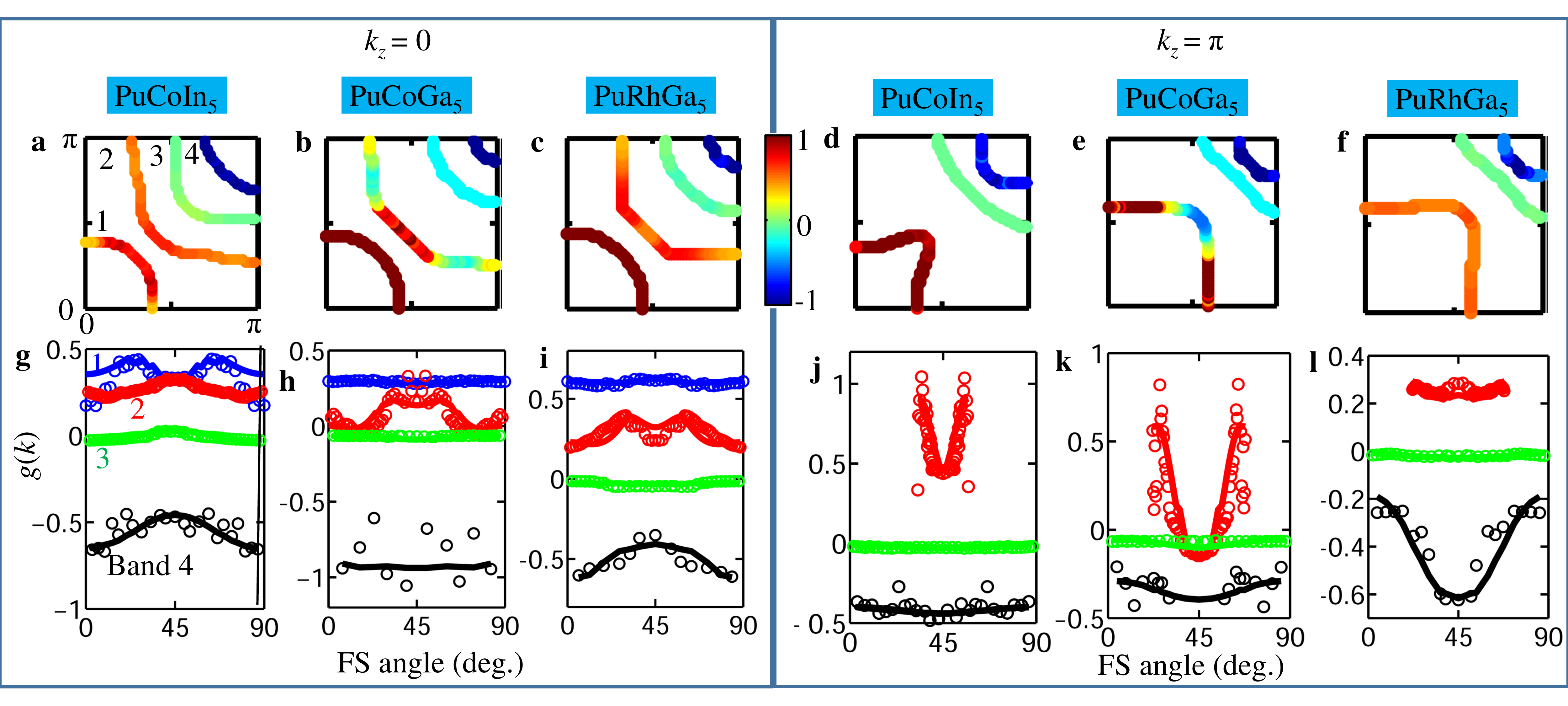}}
\vspace{16pt}
\parbox{\textwidth}{
{\bf Fig.~S3}: Top panel: Computed $k$-dependent SC gap function plotted in colormap as in the main text. Bottom panel: The same gap function of four bands plotted as a function of FS angle. The solid lines are the fit with the gap function $g(k)$ including higher harmonics as given in Eq.~(11), and the relevant parameters are given in Table~I. The FS angle is defined to be zero along the zone boundary direction (100) for band 1, and (010) for all other bands, and 45$^o$ along the diagonal direction (110) for all bands. }
\label{gapfnc}
\end{figure}

In the main text, we calculated the pairing eigenfunction $g({\bf k})$ by directly solving the eigenvalue problem, which is obtained by rewriting the linearized weak-coupling multiband gap equation. We also calculated the so-called  pairing strength $\lambda$ in two different, yet equivalent, procedures by using Eqs.~(3) and (4). We calculated the maximum eigenvalue $\lambda$ and corresponding eigenfunction $g({\bf k})$ by solving the eigenvalue matrix problem for the spin-fluctuation pairing vertex, in Eq.~(3). This involves summation over the 3D FSs of all 4 bands. For each band we expand the  $\Gamma_{nm}$ matrix into discretized Fermi momenta ${\bf k}_n$ and ${\bf k}_m^{\prime}$. In our case, since the $\Gamma_{nm}$ matrix is defined in the band basis, each matrix index $n$ assumes dimension $N_n$ (number of points for the $n^{th}$-band), with a total number of $N=\sum_{n=1}^4 N_n$ FS momenta, where $N_n$ is the number of 3D Fermi momenta for the $n^{\rm th}$ band. Therefore the matrix we diagonalize has a dimension of $N\times N$. 

The  maximum eigenvalue is proportional to the highest superconducting transition temperature and the corresponding
eigenvector gives the leading pairing function $g({\bf k})$, which is plotted in Fig.~3 (left-hand side) and in Fig~S3. The first point to notice in Fig.~S3 is that there is no clear fourfold symmetry breaking in $g({\bf k})$, and that there is no gap node along the diagonal direction. This finding clearly excludes the presence of any significant $d$-wave pairing component. 
In a final step, we analyze the pairing function in more detail by fitting it up to third harmonics of $s^{\pm}$-pairing symmetry \cite{Graser}:
\begin{equation}
g({\bf k}) = a_1(\cos{k_x}+\cos{k_y}) + 2a_2\cos{2k_x}\cos{2k_y} + 2a_3\cos{4k_x}\cos{4k_y} + a_0.
\label{eq:gapfnc}
\end{equation}
The corresponding fits are shown in Fig.~S3. The general conclusion drawn from these fits is that the coefficients for second and third harmonics $a_2$ and $a_3$, respectively, are almost an order of magnitude lower than the first harmonic of $s^{\pm}$- pairing. We also notice a weak $k_z$ dependence in the fit parameters. Furthermore, the result shows that the gap anisotropy is largest in band 2. This is where the nodes are located. The gap amplitude is smallest in band 3 and then it increases from band 2 to band 1 to band 4. This is expected from the $s^{\pm}$-pairing symmetry as the gap maxima lie at the $\Gamma$ and M points, with opposite sign. Therefore, our conclusion about the $s^{\pm}$ pairing symmetry in the Pu-based superconductor is a robust feature.

Secondly, we calculated the pairing strength through the usual projection of the eigenvalue Eq.~(3) onto selected orthogonal pairing functions with characteristic symmetry. The projected pairing strength for a given pairing symmetry $g_{\alpha}$ is calculated from Eq.~(4) of the main text, which was used earlier in Refs.~\cite{scalapino86,Graser}. Finally, the total pairing strength is obtained by summing over all indices, $\lambda^\alpha = \sum_{n,m} \lambda_{nm}^\alpha$, and is plotted as a function of the Coulomb potentials $U$ and $V$ in Fig.~3 (right-hand side). The line integrals over each FS sheet were performed over FS pockets in each corresponding $k_z$ plane, and then summed over $k_z$ slices. The intra- and interband pairing strength $\lambda_{nm}^\alpha$ is plotted in Fig.~4 as a function of the $k_z$ slices. 

\begin{table}[tb]
\parbox{\textwidth}{{\bf Table.~I}: Coefficients of various harmonics of the gap function given in Eq.~(11).}
\centering
\begin{tabular}{|c|c|c|c|c|c|c|c|c|}
\hline \hline
&\multicolumn{2}{|c|}{Band 1}& \multicolumn{2}{|c|}{Band 2}&\multicolumn{2}{|c|}{Band 3}&\multicolumn{2}{|c|}{Band 4}\\
\hline
PuCoIn$_5$ & $k_z=0$ & $k_z=\pi$ & $k_z=0$ & $k_z=\pi$ & $k_z=0$ & $k_z=\pi$ & $k_z=0$ & $k_z=\pi$\\
\hline
$a_1$ & 0.35 & -- & 0.15 & 0.5 & 0.12 & 0.12 & 0.3 & 0.2 \\
$a_2$ & 0.06 & -- & 0.025 & 0.05 & 0.015 & 0.015 & -0.06 & -0.02 \\
$a_3$ & 0.1 & -- & 0 & 0.05 & 0 & 0 & -0.06 & -0.02 \\
$a_0$ & 0.9 & -- & 0.25 & 0.21 & 0.1 & 0.1 & 0.1 & -0.08\\
\hline
PuCoGa$_5$ & \multicolumn{2}{|c|}{}& \multicolumn{2}{|c|}{}&\multicolumn{2}{|c|}{}&\multicolumn{2}{|c|}{} \\
\hline
$a_1$ & 0.13 & -- & 0.35 & 0.7 & 0.1 & 0.1 & 0.25 & 0.25 \\
$a_2$ & -0.013 & -- & -0.02 & -0.07 & -0.01 & -0.01 & -0.025 & -0.025\\
$a_3$ & 0 & -- & -0.02 & -0.07 & -0.01 & -0.01 & -0.05 & 0.05\\
$a_0$ & 0.15 & -- & 0.05 & 0.1 & 0.08 & 0.08 & -0.4 & 0.1 \\
\hline
PuRhGa$_5$ & \multicolumn{2}{|c|}{}& \multicolumn{2}{|c|}{}&\multicolumn{2}{|c|}{}&\multicolumn{2}{|c|}{} \\
\hline
$a_1$ & 0.18 & -- & 0.35 & 0.1 & 0.1 & 0.05 & 0.2 & 0.17 \\
$a_2$ & -0.01 & -- & -0.087 & 0.07 & -0.03 & 0 & -0.05 & -0.17\\
$a_3$ & -0.013 & -- & -0.025 & 0.025 & 0 & 0 & 0.2 & -0.2 \\
$a_0$ & 0.8 & -- & 0.27 & 0.24 & 0.11 & 0.05 & 0.05 & 0 \\
\hline \hline
\end{tabular}
\end{table}

\end{document}